# ATOMIC FORCE MICROSCOPY OF TRANSFORMATION TOUGHENING IN CERIA STABILIZED ZIRCONIA

Sylvain DEVILLE, Hassan EL ATTAOUI, Jérôme CHEVALIER[#]

National Institute of Applied Science, Materials Department, Associate Research Unit 5510 (GEMPPM-INSA)

Bat B. Pascal, 20 avenue Albert Einstein, 69621 Villeurbanne Cedex, France

[#] corresponding author: e-mail: jerome.chevalier@insa-lyon.fr, Tel: +33 4 72 42 61 25, Fax: +33 4 72 43 85 28

**Abstract**

We demonstrate in this paper that atomic force microscopy can be successfully used to gain further insights into the understanding of transformation toughening in ceria stabilized zirconia. Transformation was induced by stresses accumulated in the region surrounding propagating cracks in double torsion samples. The resolution provided by AFM at the surface of the samples made it possible to observe the formation of self accommodated martensite pairs in the near crack areas. The potential for transformation is found to decrease with increasing alloying addition, and is totally suppressed for 16 mol.% $CeO_2$-TZP samples. A statistical analysis of the martensite pair orientation is performed, and the relationship with the applied stress and strain fields is discussed. The contribution to transformation toughening by transformation-induced plasticity occurring in the formation of martensitic variant pairs with small net shear is demonstrated. The influence of alloying addition content on the potential for transformation toughening and fracture toughness values is finally discussed.

*Keywords* : Atomic force microscopy, toughening, (C), $CeO_2$-$ZrO_2$ (D)

## I.  INTRODUCTION

The discovery by Garvie *et al.* [1] of transformation toughening of zirconia opened the way towards a very large field of investigations for materials scientists and engineers. The potentiality for obtaining very high toughness materials by careful control of the zirconia ceramics microstructure relies on the metastable retention of the tetragonal phase at ambient temperature [2]. Upon the action of external stresses, such as in the surrounding zones of a propagating crack, tetragonal grains may transform to their stable monoclinic structure [3]. Since the transformation is accompanied by a large shear (0.16) and volume expansion (0.04), the stresses and strains induced by the transformation lead to the formation of a zone with large compressive stresses that can partially close the crack and slow down its propagation, increasing the material toughness. This phenomenon has been the object of numerous studies over the last 30 years. The martensitic nature of the t-m transformation has been investigated by various methods among which are X-Ray diffraction [4], scanning electron microscopy, optical microscopy with Normarsky contrast [5], neutron powder diffraction [6], transmission electron microscopy [7], and more recently atomic force microscopy [8-10].

Several theories have been developed to describe and predict transformation toughening [11-16]. They are based mainly on mechanical or energetic considerations. Independently of theory, it can be shown that the martensitic transformation temperature $M_s$ can be reduced by alloy additions, so that spontaneous transformation upon cooling to room temperature does not occur. The net driving force of the transformation can then be lowered even down to room temperature, until such point as an external stress is applied to the system. This is the origin of transformation toughening. Several oxides are well known to retain zirconia in its tetragonal structure at ambient temperature, totally or partially, i.e. yttria ($Y_2O_3$), ceria ($CeO_2$) or magnesia (MgO). A great number of studies have been dedicated to these three types of materials. For a review on the subject, see the work of Green [2] or Hannink et al. [17].

Several reinforcing effects might account for an increase of material toughness. The critical stress intensity factor can be described by the combination of the *matrix intrinsic toughness* and the addition of *crack shielding mechanisms*, among which transformation toughening and crack bridging arise in the particular case of ceria-doped zirconia. The prediction of the toughness can be achieved by the prediction and quantification of these different crack shielding mechanisms. In particular, the development of a reliable theory of transformation toughening requires a deep understanding of the stress and strain field distribution in the crack tip surrounding zone. The relevance of the phenomenological theory of martensitic crystallography (PTMC) [18,19] to describe the strain field is now recognized. A contribution to transformation toughening by transformation-induced plasticity results from the formation of martensitic variant pairs with large associated shear strain, absorbing some energy in the formation of these variants, energy that would otherwise be available for crack propagation, increasing thus the toughness of the material. Using the PTMC to describe transformation toughening is very appealing indeed. However, even if the theory can predict precisely the local strain distribution, achieving the comparison of theoretical calculations and experimental results has not yet been possible, as a result of the observational difficulties at the scale at which the transformation is occurring (a few nanometers). Fortunately, the development of atomic force microscopy provides a tool for investigating local relief of a few nanometers height. The potentiality to observe autoclave ageing induced martensitic relief in yttria stabilized zirconia with great precision has been demonstrated recently [10]. The aim of this study is to show that further insights can be gained from AFM experiments in the description and subsequent understanding of transformation toughening in zirconia.

II.     MATERIALS AND METHODS

*Processing*

Ceria stabilized zirconia ($CeO_2$-TZP) materials were processed by a classical processing route, using Zirconia Sales Ltd powders, with uniaxial pressing, cold isostatic pressing and sintering at 1550°C for two hours. Residual porosity was negligible. Different compositions have been processed, with increasing stabilizer content, i.e. 10, 12 and 16 mol.% $CeO_2$. Grain size (measured by the linear intercept method on thermally etched samples) and fracture toughness (measured by double torsion experiments) are given in Table 1. This shows that the grain size is the same for all the samples, the only difference lying in the alloying content. It is widely documented from the literature [2] that the larger the $CeO_2$ content, the greater the toughness.

*Double torsion tests*

The double torsion test was used to induce stress-assisted phase transformation in the surrounding of the propagating crack and to assess quantitatively transformation toughening effects. The details of the method may be found elsewhere [20, 21]. No guiding groove was machined in the specimen in order to avoid any residual stress intensity factor. A notch was machined with a diamond saw and an indentation was performed at low load (10 kg) to initiate a small crack, as seen in Fig. 1. Crack rates versus $K_I$ curves were used to determine the fracture toughness values of the materials. These curves will be discussed in another paper.

*Atomic force microscopy and optical observations*

AFM experiments were carried out with a D3100 nanoscope from Digital Instruments Inc., using oxide sharpened silicon nitride probes in contact mode, with an average scanning speed of 10 $\mu m.s^{-1}$. Since the t-m phase transformation is accompanied by large strains (4 % vol. and 16 % shear), surface relief is modified by the formation of monoclinic phase. The lateral (2 nm) and vertical (0.1 nm) resolution of AFM makes it possible to follow very precisely the transformation induced relief at the surface. The transformation zones were also photographed with an optical microscope using the Normarsky interference contrast technique (Zeiss Axiophot, Germany).

## III. RESULTS

*Transformation bands*

The surface of double torsion samples after partial crack propagation observed by optical microscopy in Normarsky contrast is shown in Fig. 1. Great differences in behavior are observed when the alloying content is increased. For low stabilizer content (10 mol.%), the formation of elongated transformed zones ahead of the crack tip is clearly observed. The presence and shape of these zones have been the object of numerous studies in the past [22-25], and their presence is thought to be related to the autocatalytic behavior of the transformation propagation of these materials. Not only the material is transformed in the surrounding zones of the crack, but some finger-like transformed bands are also found on both sides of the crack. These bands will later be referred to as *secondary bands*. All the following AFM observations were performed in particular zones of these bands, as indicated in Fig. 1a. It is already worth mentioning that the transformed bands may extend very far away from the crack tip, demonstrating thus the very high propensity for stress induced transformation of this particular composition. The formation mechanism of these bands will be discussed later.

When the stabilizer content is increased, the secondary bands disappeared, and the transformation around the crack becomes hardly visible with an optical microscope. No differences are optically observed between the 12CeTZP and 16CeTZP samples.

A detailed part of a secondary transformed band observed by AFM is shown in Fig. 2. Slight grain pop-out induced by the polishing process is visible at the surface, and the transformed band running through the micrograph is visible. A typical feature of the relief is extracted in Fig. 3, where all the martensitic characteristic features previously described for thermal martensite [10], are also visible. The presence of large shear planes is observed, planes acting together to form self-accommodating martensitic variant pairs. The formation of smaller variant pairs to accommodate strain near the grain boundaries is observed. Another zone extracted from the near crack tip zone is shown in Fig. 4. The same type of martensitic relief is observed, suggesting the near crack tip zones and secondary transformed bands are formed by the same mechanism, i.e. stress induced transformation.

### Near crack *transformation*

AFM observations of the surroundings of a propagated crack in 12CeTZP and 16CeTZP are shown in Fig. 5 and 6. Some detailed zones are highlighted in Fig. 5, where the martensitic relief is further investigated. The formation of self accommodating variant pairs is also observed, with arrows indicating the junction plane of such pairs. It is obvious that very few grains are transformed along the crack path, as opposed to what was observed for the 10CeTZP sample. Only some of the grains adjacent to the crack were able to transform under stress. This is a clear demonstration of the variation of propensity for transformation with the alloying addition modification. This point will be further discussed later.

The transformation zone width (measured at the same distance from the notch tip for all the samples) is much decreased when stabilizer content is increased, and no transformation at all is observed when the stabilizer content reaches 16 mol.%. Though very large stresses are expected in the surrounding of the crack, these stresses were not high enough to overcome the transformation energy barrier and trigger the transformation. The variation of transformed zone width and toughness as a function of stabilizer content is plotted in Fig. 7. It is quite clear from the graph that the toughness is directly related to the propensity for transformation.

*Transformation sequence*

The very local observations of transformation induced relief bring new information about the toughening mechanism sequence. Fig. 8 shows a transformed grain with the propagated crack running through it. A fragmentation of transformed planes due to the crack is observed. It can therefore be safely assumed that the transformation occurred *before* crack propagation. While the crack is still stationary, stresses are building up in its surroundings. Once these stresses are high enough, transformation of the grains in these zones is triggered, absorbing some of the stresses. If stresses continue increasing, the crack will be free to further propagate in the transformed zones. These results have been confirmed by complementary acoustic emission experiments [26].

Moreover, for lattice correspondence CAB ($a_t$, $b_t$, $c_t$ axis of the tetragonal phase changes into $c_m$, $a_m$, $b_m$ axis of the monoclinic phase), all the transformation strain can be accommodated vertically if the grain has its $c_t$ axis nearly perpendicular to free surface [27]. In this particular case, no residual stresses should be expected in the bulk once the grain is fully transformed. There will not be any stresses opposed to crack propagation. This can further explain the observation of the crack running straight though the transformed grain, without being deviated from its initial path.

*Relationships with stress field*

Among the inputs required by transformation toughening theories [11-15], the nature and the magnitude of strain fields in the surrounding zones of the crack tip are of prime importance. Since the precise determination of these fields was not possible up to now [17], predictions relied only on the calculations results. Almost all of the calculations developed so far are based on the Eshelby formalism [27], describing strains induced by the formation of the monoclinic products of the reaction in the tetragonal matrix. Further progress has then been made by using the PTMC, but the lack of comparison with experimental evidence was still a great limitation of further improvement of the theories. The scale at which the relief can be described by AFM (e.g. see Fig. 2) is a great step toward a deeper understanding of the transformation mechanism and validation of the developed theories. In particular, the orientation relationship of the observed relief with the applied stress is worth further analysis. Based on the representation described in Fig. 9, a statistical analysis of the orientation deviation of the variant pairs in the 10CeTZP sample was performed. The orientation of 130 variant pairs was measured, to get a statistically significant average orientation. The orientation of each pair was measured with respect to the crack propagation direction. The distribution of the orientation deviation is plotted in Fig. 10. An average value of 27° was obtained, while the secondary band orientation was found to be 26°, which means all the analyzed transformed variants are lying in the direction of the transformed band propagation. Some of the grains having their $c_t$ axis close to the free surface normal and with a potential junction plane lying in a *perpendicular* position to the crack propagation direction were preferentially activated (see for example Fig. 4). Less transformed grains are found when their junction plane is deviating from the crack direction. The same analysis was performed on the variants in the near surroundings of the propagated crack, and an average value of 4° was found. The behavior is thus the same in the secondary bands and in the *near crack* transformed zones.

As far as the propensity for transformation is concerned, two crystallographic features must be considered to explain the observed behavior, i.e. the potential junction plane orientation with respect to the *stress field* and the orientation relationship of the potential junction plane with the *sample surface*. If the junction plane is perpendicular to the surface, as for example in Fig. 3 the large residual shear can potentially accommodate all of the strain induced by the transformation in the vertical direction [28]. This way of accommodating the strain is much more favorable than when the junction plane is lying parallel to the surface, which would mean the strain should be accommodated in lateral directions, which is much more difficult, considering the restricting mechanical effect of the matrix. The propensity for transformation is therefore a consequence of the combination of these two parameters, though they do not have the same relative importance. Other factors such as local residual stress inhomogeneities and microstructural defects may also play a role, though of second order; their influence is consequently not discussed here. Comparing this analysis with the observed relief of Fig. 2 it seems that the controlling factor is the relationship of the junction plane and the applied stress more than the orientation to the surface, since among the transformed grains, just a few present a junction plane perpendicular to the surface.

A uniaxial stress state seems to be the more favorable state for the transformation. The possibility of inducing the transformation by uniaxial compression was first reported by Lankford [29]. This particular point may be understood by considering the fact that all of the transformed grains present self accommodating variant pairs. The net shear induced by the transformation is therefore systematically annihilated upon transformation completion. In the case where self accommodation would not be occurring (leading to a large net shear), strain induced transformation should be accommodated only by volume increase. In this case, applying a compressive stress state would inhibit the transformation. When transformation induced strain is accommodated more by shear than volume increase, either uniaxial tensile or compressive stress will be favorable and shear stress prevails over volume increase (opposed to the transformation) in activating the transformation.

## IV.     GENERAL DISCUSSION

The new features brought by this study provide valuable information in regards to the transformation toughening theories requirements and inputs. The first point is the influence of alloying addition on the potential for transformation toughening. For the 10CeTZP samples, a very large propensity for martensitic transformation is found. The Ms temperature has been reduced close to room temperature, so that transformation can be easily stress induced, providing a large potential for transformation toughening. The toughness measured by double torsion was indeed found to be very high, i.e. 18 MPa.m-1/2. On the other hand, when the transformation is less easily stress induced, i.e. 12 mol.% $CeO_2$ samples, the toughness falls down to 8 MPa.m-1/2. The transformation toughening contribution to toughness becomes negligible, as observed on the AFM micrographs. When the alloying content reaches 16 mol.%, no transformation at all is observed, and the toughness falls down to 4.3 MPa.m-1/2. The only remaining crack shielding mechanism is crack bridging. It is also worth noticing that since the grain size was the same in all the samples, the magnitude of crack bridging can be assumed to be the same for all the samples, so that it does not interfere with the present observations.

The observed autocatalytic transformed bands ahead of the crack tip can be explained by the PTMC. In fact, the transformation induced shear strain can be accommodated by the formation of self accommodating martensitic variant pairs. However, residual stress not accommodated by the transformation-induced plasticity may be used to trigger the transformation of neighboring grains, providing the orientation relationships of the two grains are energetically favorable for the transformation to proceed, as shown by previous studies [10]. The formation of these elongated transformed zones could thus be explained by the autocatalytic nature of the transformation of ceria-stabilized zirconia.

More important was the observation that transformed grains were always formed by self accommodating variant pairs, implying the presence of large residual local shear after transformation completion but *very low net shear*, if any. This may actually have major implications as far as the transformation toughening theories are concerned. In fact, all the *local* investigations of the last twenty years were performed by transmission electron microscopy, on thin foils samples. The microstructural environment of the samples is modified, in particular during the grinding process, and the stress state may be modified by the very low thickness of the foils. The comparison of such results with macroscopic observations on bulk samples may therefore be questioned. In particular, all the transformation toughening models developed so far demonstrate a net improvement to transformation toughening when a net shear component is added to the transformation strains [14,15,30]. In the same time, it was believed [31] that only transforming zirconia particles in stable matrices (e.g. Mg-PSZ, zirconia toughened alumina, etc…) would exhibit self accommodating variant pair formation, leading to a reduced net shear, due to the restricting influence of the matrix. These twin-related variants were also thought to be a much rarer occurrence in TZP materials. It is however clearly demonstrated here that nearly all the transformed grains of Ce-TZP exhibit twin related variants, leading to very limited net shear. According to Hannink [17], "more potent toughening is expected in YTZP than in PSZ or ZTC, [which is] contrary to all the available experimental evidences". Considering the evidences provided here, it is quite clear that differences between TZP and PSZ or ZTC are not lying in differences in the net shear of the transformation, since all these types of materials exhibit twin related variants and thus great accommodation of the shear strain. Most of the difficulties arising from the confrontation of transformation toughening theories and experimental evidences may therefore be elucidated.

Finally, it is worth mentioning that incorporating the above measured transformed zone width in any quantitative transformation toughening model would very likely lead to mistaken predicted values. The transformation zones in the bulk are likely to be very different from that at the surface. AFM observations only provide observation of *surface* transformation. Different techniques must be used to investigate the transformation zone shape in the bulk.

## V. CONCLUSIONS

The transformation toughening behavior of 10, 12 and 16 mol.% $CeO_2$-TZP was investigated by atomic force microscopy on double torsion sample surfaces. Elongated autocatalytic transformed zones were observed for 10 mol.% ceria samples ahead of the crack tip and also almost parallel to the crack propagation direction. AFM allowed direct observation of martensitic relief in the surrounding zones of propagated cracks. Transformation toughening was found to decrease with alloying addition, with the apparent transformation zone width at the surface being directly related to the toughness and alloying content. No transformation was observed for samples with 16 mol.% ceria addition.

The orientation of the martensitic variants with respect to the crack propagation direction and stress field was analyzed statistically and found to be in good agreement with previous experimental results. In particular, it was shown that the transformation could be induced by a uniaxial stress state. More importantly, the contribution to transformation toughening by transformation-induced plasticity occurring in the formation of martensitic variant pairs with large associated local shear and small net shear was demonstrated. The systematic formation of twin related variants in TZP materials can explain the discrepancies observed up to now between the theories of transformation toughening and the experimental evidence, since the net shear is the same in tetragonal zirconia polycrystals, partially stabilized zirconia or zirconia toughened composites materials. AFM appeared thus as an extremely powerful tool to investigate the transformation toughening mechanism in zirconia based ceramics.

ACKNOWLEDGEMENTS

Authors are grateful to the CLAMS (Consortium de Laboratoires d'Analyse par Microscopie à Sonde locale) for using the nanoscope. The authors would also like to acknowledge the European Union for the financial support under the GROWTH2000, project BIOKER, reference GRD2-2000-25039 and the Rhône-Alpes region for financial support under MIRA project.

| Material | Ceria content | Sintering temperature | Grain size (linear intercept) | Fracture toughness |
|---|---|---|---|---|
| 10Ce-TZP | 10 mol.% | 1550°C | 3.7 µm | 18 Mpa.m$^{-1/2}$ |
| 12Ce-TZP | 12 mol.% | 1550°C | 3.5 µm | 8.1 Mpa.m$^{-1/2}$ |
| 16Ce-TZP | 16 mol.% | 1550°C | 3.4 µm | 4.3 Mpa.m$^{-1/2}$ |

Table 1: Materials of the study. All the samples exhibit a similar grain size. The only variable is the stabilizer content. Fracture toughness values were provided by double torsion relaxation experiments.

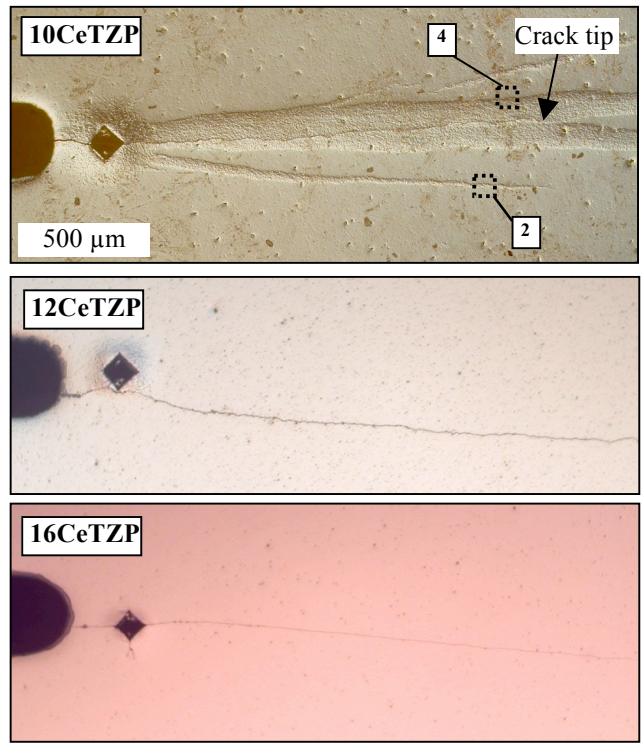

Fig. 1: Optical observation of a partially propagated crack at the surface of the various double torsion samples. Some finger-like elongated transformed bands could be seen around the crack of the 10Ce-TZP sample. The AFM observation zones are indicated on the micrograph. Arrow indicates the crack tip.

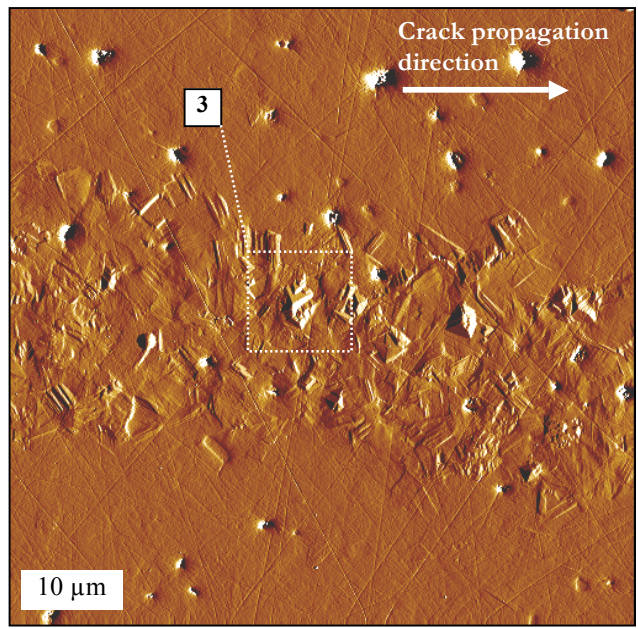

Fig. 2: AFM observation of a transformed band in 10CeTZP. Some grain pop-out induced by polishing could be seen. The transformed band exhibiting typical martensitic relief is running through the entire micrograph.

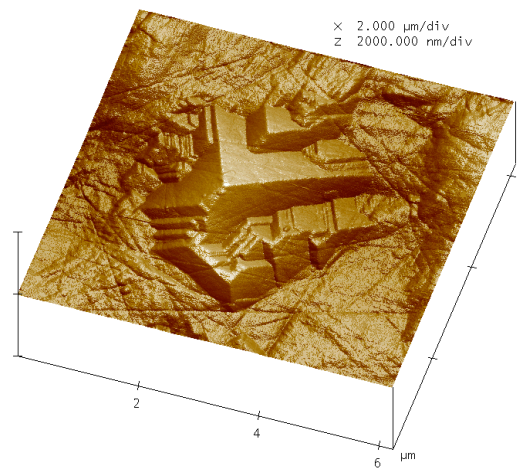

Fig. 3: Detailed zone of Fig. 2 (10CeTZP) showing a typical stack of self-accommodating martensitic variant pairs. Note the very large shear strain induced by the transformation.

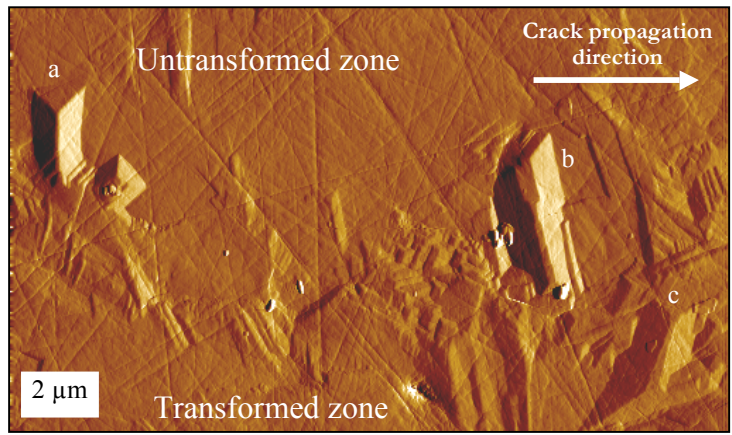

Fig. 4: Border zone of the surrounding of the propagated crack in 10Ce-TZP. Transformed variants (a-c) perpendicular to the crack path are clearly visible, for grains having their $c_t$ axis nearly perpendicular to the surface.

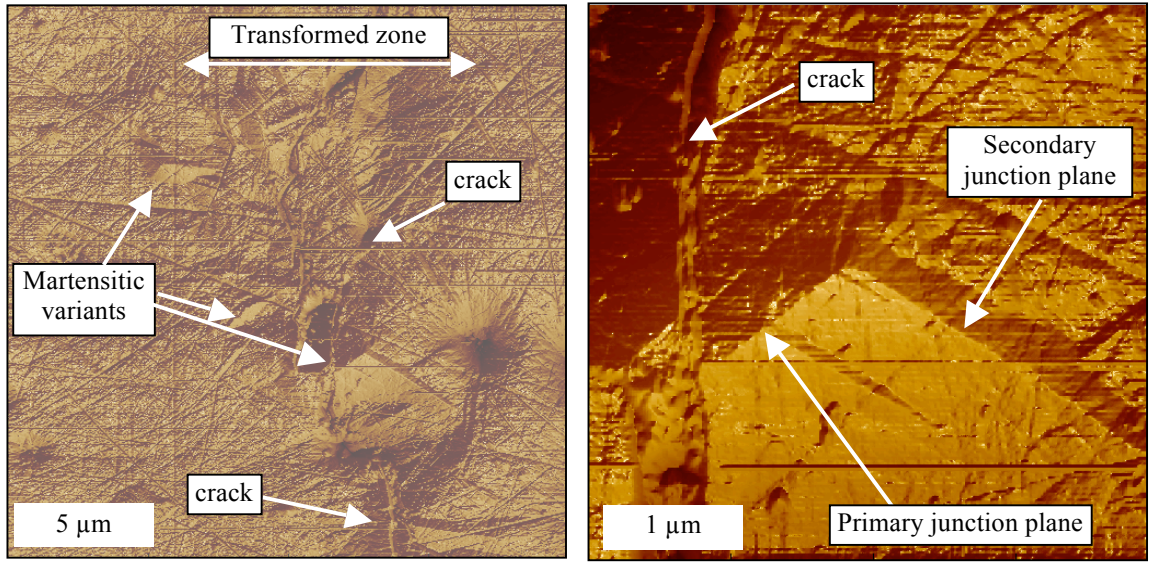

Fig. 5: Surrounding of the propagated crack in 12Ce-TZP. Transformed variants are clearly visible. The transformed zone width is much smaller than for the 10Ce-TZP sample.

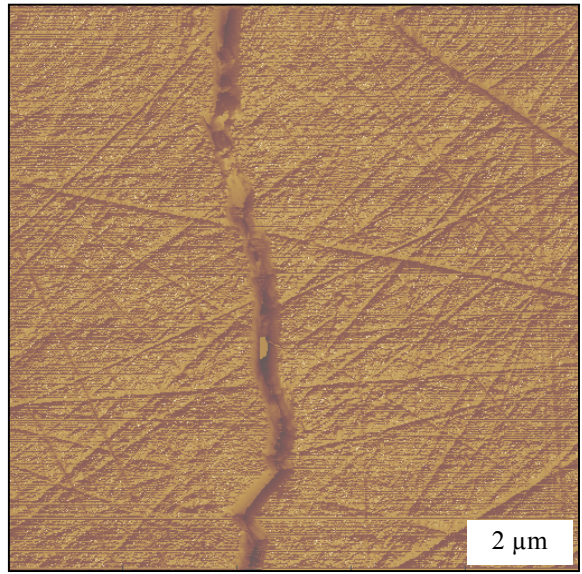

Fig. 6: Surrounding of the propagated crack in 16Ce-TZP. No transformation at all is observed. Residual scratches from polishing are observed.

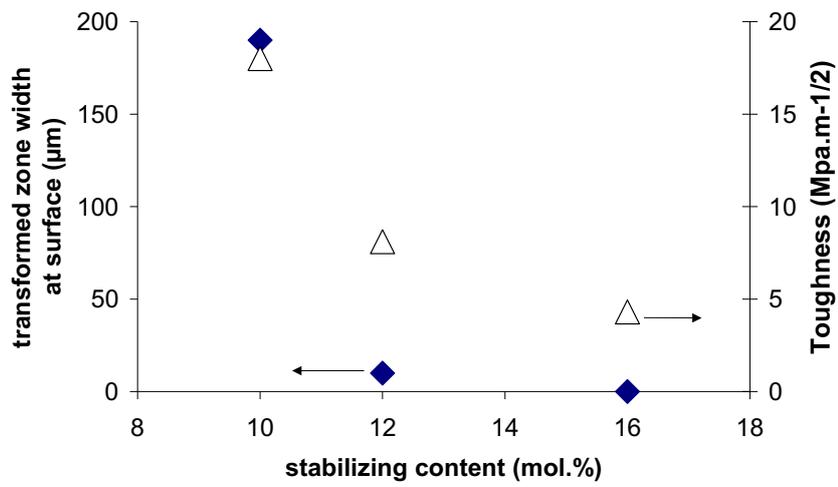

Fig. 7: Transformed zone width at surface and toughness as a function of alloying content. The toughness is directly related to the width of the transformation zone.

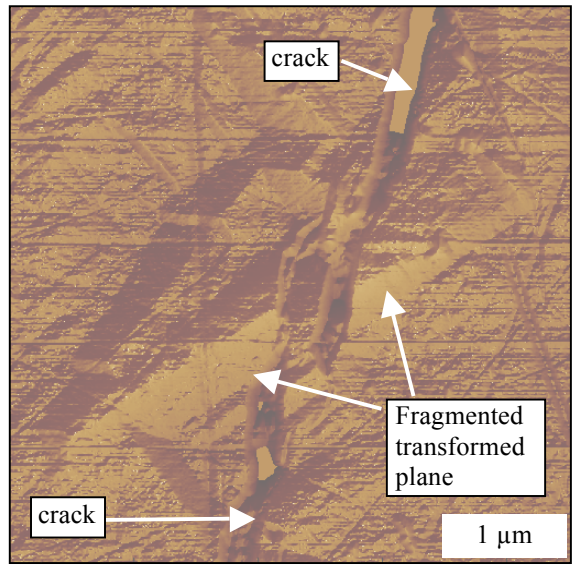

Fig. 8: Surrounding of the propagated crack in 12Ce-TZP. The grain was transformed before crack propagation, and the transformed plane were fragmented when the crack ran through it. No residual stresses are expected when the transformation strain is accommodated vertically, so that it was possible for the crack going through the transformed grain instead of avoiding it.

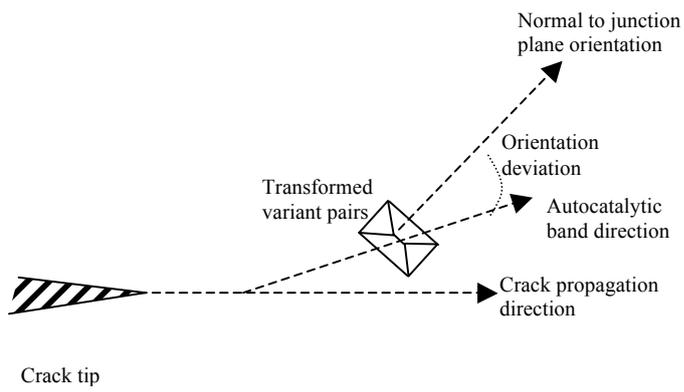

Fig. 9: Calculation of the orientation deviation of the transformed grain with the crack.

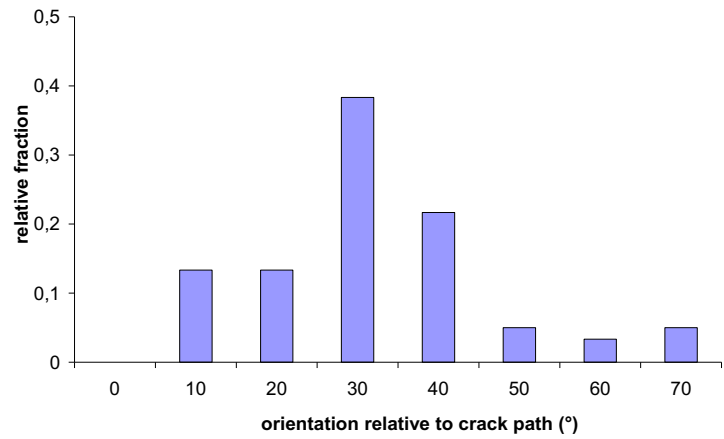

Fig. 10: Orientation deviation distribution (see text for details). A preferential orientation of the junction planes (26° to crack path) perpendicular to band direction (27° to crack path) is observed, suggesting a strong dependence of the grains sensitivity to transformation to the crack path orientation.